\documentclass[aps,prl,twocolumn,showpacs]{revtex4}
\usepackage{pifont}
\usepackage{stmaryrd}
\usepackage{bbm}
\usepackage{amssymb}
\usepackage{amsmath}
\usepackage{graphicx}
\usepackage{txfonts}

\begin{document}
\title{Experimental Non-Local Generation of Entanglement from
Independent Sources\footnote{Supported by the Marie Curie Excellence
Grant of the EU, Chinese Academy of Sciences, and the National Basic
Research Program under Grant No 2006CB921900.\\

$^{**}$Email: jinxm@ustc.edu.cn}}

\author{Xian-Min Jin$^{1,2**}$, J\"ugen R\"och$^2$, Juan  Yin$^1$, Tao Yang$^1$}

\affiliation{$^1$Hefei National Laboratory for Physical Sciences at
Microscale and Department of Modern Physics, University of Science
and Technology of China, Hefei 230026 \\
$^2$Physikalisches Institut, Ruprecht-Karls-Universit\"{a}t
Heidelberg, Philosophenweg 12, 69120 Heidelberg, Germany}

\begin{abstract}
We experimentally demonstrate a non-local generation of entanglement
from two independent photonic sources in an ancilla-free process .
Two bosons (photons) are entangled in polarization space by steering
into a novel interferometer setup, in which they have never meet
each other. The entangled photons are delivered to polarization
analyzers in different sites, respectively, and a non-local
interaction is observed. Entanglement is further verified by the way
of the measured violation of a CHSH type Bell's inequality with
S-values of 2.54 and 27 standard deviations. Our results will shine
a new light into the understanding on how quantum mechanics works,
have possible philosophic consequences on the one hand and provide
an essential element for quantum information processing on the other
hand. Potential applications of our results are briefly discussed.
\end{abstract}

\pacs{03.65.Ud, 03.67.Mn, 42.50.Dv} \maketitle

Entanglement is considered to be one of the most profound features
of quantum mechanics and is extremely important in quantum
information. It lies at the heart of the Einstein--Podolsky--Rosen
paradox,\cite{EPR1935} Bell's inequalities,\cite{Bell1964} and the
nonlocality of quantum mechanics, and has comprehensive application
in quantum communication\cite{Gisin2002,Duan2001} and quantum
computation.\cite{Nielsen2002} So far, several methods to generate
entanglement have been proposed and realized, as illustrated in
Figs.\,1(a) and 1(b). One way is to entangle two particles which
root in a common source,\cite{kwiat1995,Freedman1972,Rarity1990}
such as from the process of spontaneous parametric down conversion
(SPDC) in a nonlinear optical crystal, where a ultraviolet pump
photon decays with low probability into two infrared photons in
terms of conservation of energy and momentum (Fig. 1(a)). The other
way is to have two particles interacting with each
other,\cite{Shih1988} typically, oppositely polarized photons, say
horizontal (H) and vertical (V), impinging on a beamsplitter (BS)
from opposite input ports, photons can be entangled by exchange
interaction and postselecting photons from opposite output ports,
i.e. the well-known Hong--Ou--Mandel (HOM) \cite{Hong1987}
interference (Fig. 1(b)). An otherguess idea to obtain entanglement
has been proposed \cite{Bennett1993,Zukowski1993,Bose1998} and
realized experimentally by Pan {\it et al.},\cite{Pan98}  i.e. the
so-called entanglement swapping (Fig. 1(c)), which entangles freely
propagating particles that never physically interacted with one
another or which have never been dynamically coupled by any other
means. As is shown in Fig. 1(c), two simultaneously produced
entangled photons, pair EP1--EP2 and pair EP3--EP4. One photon from
each pair (photon EP2 and EP3) is subjected to a Bell-state
measurement (BSM) which can be realized with a BS and a coincidence
measurement between two outputs. This results in projecting the
other two outgoing photons EP1 and EP4 into an entangled state.

\begin{figure}[ptb]
\begin{center}
\includegraphics[width=8.4cm]{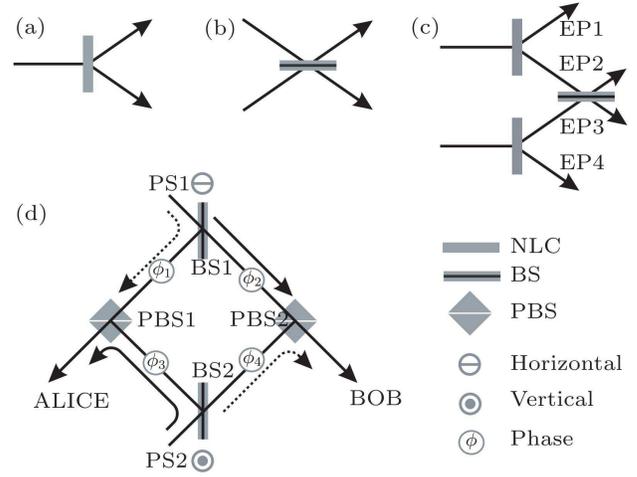}
\caption{Method of non-local generation of entanglement. (a)
Entanglement rises from the process of spontaneous parametric down
conversion by pumping a nonlinear crystal. (b) Entangle two photons
by employing HOM interference. (c) Entanglement swapping. (d) Our
scheme of non-local generation of entanglement from independent
photonic sources. The solid and dotted lines indicates the two cases
we select. NLC: nonlinear crystal; BS: beamsplitter; PBS: polarized
beamsplitter; EP1,EP2,EP3,EP4: entangled photons 1,2,3,4; PS1,PS2:
independent photonic sources 1, 2.} \label{entanglement}
\end{center}
\end{figure}

A marvelous scheme proposed by Yurke {\it et al.}\cite{Yurke1992}
shows that a kind of Bell-type inequality violation occurs even
between two independent particle sources, and an interference
between two indistinguishable electrons from independent sources has
been realized by Neder {\it et al.}\cite{Neder2007} In this Letter,
we experimentally demonstrate a non-local generation of entanglement
from independent photonic sources, which is totally different from
the means mentioned above. By using a novel apparatus
\cite{Boschi2007,Yurke1992} and post-selection technology we obtain
an entangled state from two independent photons without coming from
a common source, interacting in the past or ancillary entangled
states.

A diagram of our scheme is shown in Fig. 1(d), two independent,
separated, photonic sources PS1 and PS2 with polarization H and V
are delivered into input ports of the 50:50 beamsplitters BS1 and
BS2, respectively. The other input ports of BS1 and BS2 serve to
vacuum. After the action of the BS1 and BS2, the single photons from
PS1 and PS2 will be split coherently to two spatial modes of each BS
in a superposition style as:
\begin{eqnarray}
&&\left\vert \psi\right\rangle_1=\left\vert H\right\rangle
_{1}\varotimes\frac{1}{\sqrt{2}}
\big(iA^{\dagger}_{1}+B^{\dagger}_{1}\big)\left\vert 0
\right\rangle , \notag \\
&&\left\vert \psi\right\rangle_{2}=\left\vert V\right\rangle
_{2}\varotimes\frac{1}{\sqrt{2}}\big(iA^{\dagger}_{2}+B^{\dagger}_{2}\big)\left\vert
0 \right\rangle , \label{Eq.1}
\end{eqnarray}
where we have used the convention to express the output state of BS1
and BS2 in the particle creation operators. The factor $i$ is a
consequence of unitarity. It corresponds physically to a phase jump
upon reflection at a BS. Equation (1) describes the fact that PS1
(PS2) can be found with equal probability (50$\%$) in either of the
output modes to Alice $A_{1}$ ($A_{2}$) or to Bob $B_{1}$ ($B_{2}$).
The four output beams in mode $A_{1}, B_{1}, A_{2}, B_{2}$ will
undergo a corresponding phase shift $\phi_{1}, \phi_{2}, \phi_{3},
\phi_{4}$, then recombine and superpose at polarizing beamsplitter
(transmit H and reflect V) PBS1 and PBS2 with a partnership ($A_{1},
A_{2}$) $\Rightarrow$ A and ($B_{1}, B_{2}$) $\Rightarrow$ B.
Considering similar unitary transformation for reflection from PBS,
the state in Eq. (1) thus evolves into:
\begin{eqnarray}
&&\left\vert \psi^{'}\right\rangle_{1}=\left\vert H\right\rangle
_{1}\varotimes\frac{1}{\sqrt{2}}\big(ie^{i\phi_{1}}A^{\dagger}_{1}+e^{i\phi_{2}}B^{\dagger}_{1}\big)\left\vert
0 \right\rangle,\notag \\
&&\left\vert \psi^{'}\right\rangle_{2}=\left\vert V\right\rangle
_{2}\varotimes\frac{1}{\sqrt{2}}\big(-e^{i\phi_{3}}A^{\dagger}_{2}+ie^{i\phi_{4}}B^{\dagger}_{2}\big)\left\vert
0 \right\rangle .\label{Eq.2}
\end{eqnarray}
It should be noted that photons PS1 and PS2 are not distinguishable
anymore if they arrive at the PBS1 or PBS2 simultaneously. The
spatial parts of photons $A_{1}$ and $A_{2}$ ($B_{1}$ and $B_{2}$)
will be integrate to A (B) with the action of PBS1 (PBS2). The total
outgoing state including both the spatial and the spin part will
thus be written as
\begin{eqnarray}
&&\left\vert \psi^{'}\right\rangle_{1}\varotimes\left\vert
\psi^{'}\right\rangle_{2}\notag\\
&&=\frac{1}{\sqrt{2}}\Big(-ie^{i(\phi_{1}+\phi_{3})}\left\vert
H\right\rangle _{A}\left\vert V\right\rangle
_{A}+ie^{i(\phi_{2}+\phi_{4})}\left\vert H\right\rangle
_{B}\left\vert V\right\rangle _{B}\notag\\
&& -\big(e^{i(\phi_{1}+\phi_{4})}\left\vert H\right\rangle
_{A}\left\vert V\right\rangle
_{B}+e^{i(\phi_{2}+\phi_{3})}\left\vert V\right\rangle
_{A}\left\vert H\right\rangle _{B}\big)\Big). \label{Eq.3}
\end{eqnarray}
With the post selection technology, Alice and Bob just pick out the
events only one photon in each site. Omitting an overall phase
shift, the final two-photon state shared between Alice and Bob will
be a tunably polarization-entangled state:
\begin{eqnarray}
\left\vert \Psi\right\rangle_{AB}=\frac{1}{\sqrt{2}}\Big(\left\vert
H\right\rangle _{A}\left\vert V\right\rangle
_{B}+e^{i\phiup}\left\vert V\right\rangle _{A}\left\vert
H\right\rangle _{B}\Big),\label{Eq.4}
\end{eqnarray}
where $\phiup=\phi_{2}+\phi_{3}-\phi_{1}-\phi_{4}$. By properly
adjusting our apparatus such that $\phiup$=0 or $\pi$, the generated
state shared by Alice and Bob will be exactly maximal
polarization-entangled state.
\begin{eqnarray}
\left\vert
\Psi^{+}\right\rangle_{AB}=\frac{1}{\sqrt{2}}\Big(\left\vert
H\right\rangle _{A}\left\vert V\right\rangle _{B}+\left\vert
V\right\rangle _{A}\left\vert
H\right\rangle _{B}\Big),\notag\\
\left\vert
\Psi^{-}\right\rangle_{AB}=\frac{1}{\sqrt{2}}\Big(\left\vert
H\right\rangle _{A}\left\vert V\right\rangle _{B}-\left\vert
V\right\rangle _{A}\left\vert H\right\rangle _{B}\Big).\label{Eq.5}
\end{eqnarray}

In the process, PS1 (PS2) has the same probability $P=0.5$ to pass
the BS1 (BS2) or to be reflected. Thus, four different possibilities
arise: (1) PS1 $\Rightarrow$ A, PS2 $\Rightarrow$ A; (2) PS1
$\Rightarrow$ B, PS2 $\Rightarrow$ B; (3) PS1 $\Rightarrow$ A, PS2
$\Rightarrow$ B; (4) PS1 $\Rightarrow$ B, PS2 $\Rightarrow$ A. Each
of the four cases occurs with the same probability. The cases (1)
and (2), in which PS1 and PS2 meet with each other, are filtered by
our conditional trigger strategy, therefore have no contribution to
our finally obtained state in Eq. (5). The cases (3) and (4) are
selected by Alice and Bob, whose trajectories are shown in Fig.
1(d), with the dotted line for the case (3) and the solid line for
the case (4). In both the cases, PS1 and PS2 always fly to opposite
directions so that they never meet each other. However, the cases
(3) and (4) themselves are not differentiable and can be indicated
quantum mechanically in the superposition of these two cases, i.e.
entanglement.

\begin{figure}[ptb]
\begin{center}
\includegraphics[width=8.4cm]{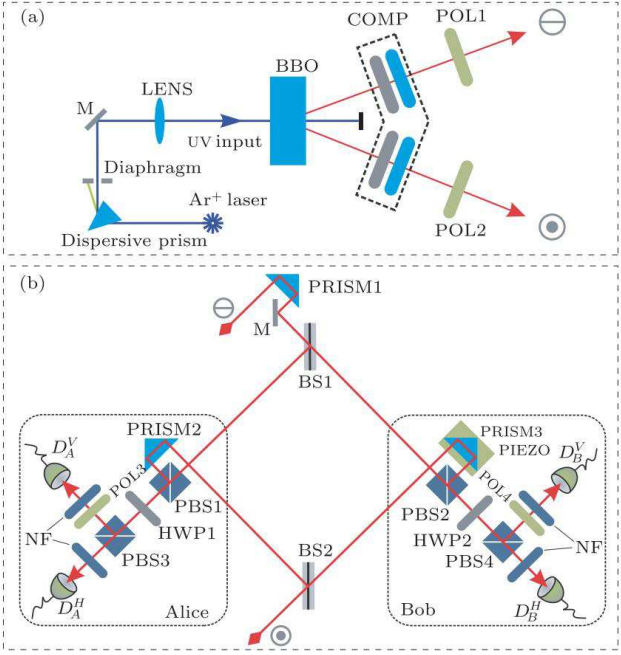}
\caption{Schematic of the experimental setup. (a) Preparation of two
photonic sources. A UV beam from argon-ion laser (351.1 nm, 300 mW)
is directed into the BBO crystal to create photon pairs with
wavelength 702.2 nm. Two compensators (COMP) are used to offset the
birefringent effect caused by the BBO crystal during parametric
down-conversion. Polarizers POL1 and POL2 setting at 0$^\circ$ and
90$^\circ$ to prepare single photon H and V in different spatial
modes. (b) The setup for nonlocally creating entanglement. The
PRISM1,PRISM2 and PRISM3 are built in micrometer-precision manual
positioning stages to balance path length for PS1(H) and PS2(V) to
PBS1 and PBS2, and an additional nanometer stepsize piezo
translation stage is mounted to PRISM2 to tune or fix $\phiup$
arbitrarily and therewith has full control over the phase. A
polarization analyzer in Alice (Bob) side comprises an HWP, a PBS
and two single-photon counting modules (Perkin-Elmer, SPCM-AQR-13
operated in Geiger-mode).)} \label{setup}
\end{center}
\end{figure}

A schematic drawing of the experimental realization and analyzing of
the non-locally generated entangled state is shown in Figs. 2(a) and
2(b). Our scheme works ideally with true independent single photon
input. In our proof-of-principle experiment, we employ disentangled
photons from SPDC sources as the two input photons. As shown in Fig.
2(a), an argon-ion UV laser beam (with a power of 300 mw, a waist of
80 $\muup$m and a central wavelength of 351.1 nm) passes through a 2
mm beta-barium-borate (BBO) crystal to generate photonic pairs
(702.2 nm) with type-I$\!$I phase matching. Unwanted
laser-fluorescence is minimized by a dispersion prism. With narrow
bandwidth interference filters (NIF) of 1.5\,nm in front of single
photon detectors, we collect about 12000 pairs of photons per
second, and about 150000 per second for single photon counts in each
side. Down-converted extraordinary and ordinary photons have
different velocities and travel along different paths inside the
crystal due to the birefringent effect of the BBO crystal. The
resulting walk-off effects are compensated by a combination of a
half wave plate ($\lambda/2, HWP$) and an additional 1\,mm BBO
crystal in each arm. In this step, by checking the entanglement
visibility between the two created photons we can obtain exactly
identical photons in frequency with the assistance of type-I$\!$I
phase matching. Further followed by a pair linear polarizers POL1
and POL2 (extinction ratio of 10000:1) setting at $0^{\circ }$ and
$90^{\circ }$, two independent photons with H and V polarization
respectively, which are located at different sites and with
accurately the same frequency, will be obtained.

As is shown in Fig. 2(b). The prepared two photons are delivered
into the novel interferometer we have illustrated above. At the
output ports of the interferometer located in Alice and Bob, the
polarization analyzers are employed to measure ingoing photons. We
use 1.5nm NIFs to increase the coherent length, define the exact
spectral mode and remove all background light. Highly
extinction-ratio polarizer Pol3 (Pol4) is used to eliminate the
imperfect reflection of PBS1 (PBS2) for Vertical polarization.

One prerequisite to observe nonlocal interaction is to ensure that
the two input photons are indistinguishable at each side. Perfect
spatial and temporal overlaps on the PBS1 and PBS2 are necessary,
which are highly related to the visibility of the created entangled
states. Experimentally, all the photons are collected with
single-mode fibers to define the exact spatial mode. All NIFs are
set in front of each detector to define the exact spectral mode.
Additional prism 1, prism 2 and prism 3 built in
micrometer-precision manual positioning stages are employed to
achieve accurate temporal overlap for two photons on PBS1 and PBS2.
To check whether the condition of arriving at PBS1 and PBS2 at the
same time for two photons is fulfilled, we exploit and apply HOM
type interference \cite{Hong1987} at the both of Alice and Bob's
sites. The first and second terms of Eq.\,(3) denote the cases that
both photons reach the same site. These rejected parts can be reused
to confirm the simultaneous arrival of two photons. Considering the
indistinguishability of the two photons, both the terms can be
rewritten in $\left\vert +\right\rangle/\left\vert -\right\rangle$
basis as $\left\vert +\right\rangle_{1}\left\vert
+\right\rangle_{2}+\left\vert -\right\rangle_{1}\left\vert
-\right\rangle_{2}$. The optical axis of HWP1 and HPW2 are rotated
to $22.5^{\circ }$, the photon with $\left\vert +\right\rangle$
polarization will fire the detectors $D^{H}_{A}$ and $D^{H}_{B}$,
the one with $\left\vert -\right\rangle$ polarization will fire
$D^{V}_{A}$ and $D^{V}_{B}$ definitely. Thus a HOM-type dip will
emerges at perfect temporal overlap if coincidence measurements of
$D^{H}_{A}$-$D^{V}_{A}$ ($D^{H}_{B}$-$D^{V}_{B}$) are performed. By
adjusting prisms 1 and 2, we have measured the two-fold coincident
counts as a function of scanning position of prisms 1 and 2.
Ideally, there should be completely destructive interference if the
wavepackets of the two photons overlap perfectly. However, it is
difficult to make the two wavepackets absolutely identical or
exactly overlapped in practice. In our experiment, we achieve a very
high visibility of the dips
$V_{A}=(C_{plat}-C_{dip})/C_{plat}=(96.6\pm0.5)\%$ for Alice and
$V_{B}=(95.9\pm 0.8)\%$ for Bob respectively, where $C_{plat}$ is
the non-correlated coincidence rate at the plateau and $C_{dip}$ is
the interfering coincidence rate at the dip, see Figs. 3(a) and
3(b). The extremely high HOM-type interference visibility ensure
considerably high entanglement visibility we can obtain.

\begin{figure}[t]
\includegraphics[width=8.4cm]{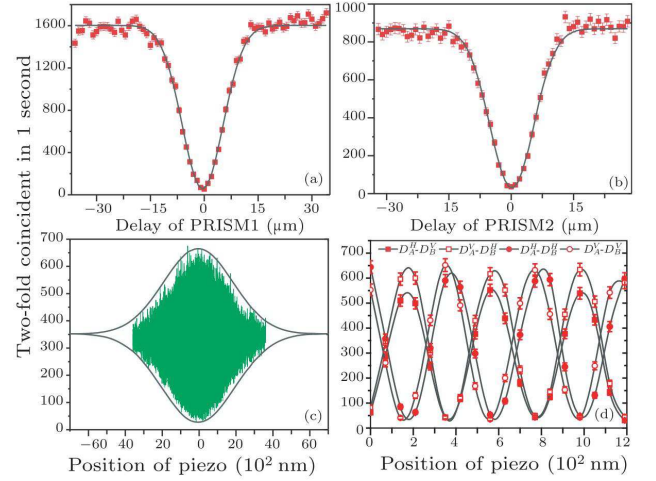}
\caption{Demonstration of the way to make two photons
indistinguishable and the non-local interference fringe. (a) and (b)
Hong--Ou--Mandel dips observed in Alice's side (a) and Bob's side
(b). (c) The envelope of the observed two-fold coincidence of
$D^{H}_{A}$-$D^{V}_{B}$. (d) We use a piezo translation stage to
move PRISM2 to perform a fine scan around the centre of the
envelope. By setting the piezo system to a position where we observe
maximum two-fold coincidence of $D^{H}_{A}$-$D^{V}_{B}$, we can
achieve $\phiup=\pi$.} \label{alldata}
\end{figure}

Once the indistinguishability of two photons are fulfilled at both
sites, we measure the two-fold coincidence between the output modes
toward detectors $D^{H}_{A}$ and $D^{V}_{B}$ with HWP1 and HWP2
setting at $22.5^{\circ }$. An interference fringe can be observed
by performing fine adjustment of the position of either PRISM1 or
PRISM2. By using a piezo translation stage (minimum step size 1nm)
to move prism 2, we perform a fine scan to measure the envelope of
the two-fold coincidence $D^{H}_{A}$-$D^{V}_{B}$ with a step size of
50\,nm, see Fig. 3(c). By setting the piezo system to a position
where we observe maximal two-fold coincidence of
$D^{H}_{A}$-$D^{V}_{B}$ as shown in Fig. 3(d), we can achieve
$\phiup=\pi$ and obtain the $\left\vert \Psi^{-}\right\rangle_{AB}$
state.

The two-photon entangled state shared by Alice and Bob can be
verified by a Clauser--Horne--Shimony--Holt (CHSH) type
inequality,\cite{chsh1969} which is one type of the Bell
inequalities. The polarization correlation function is defined as
follows:
\begin{equation}
E(\theta_A,
\theta_B)=\frac{N_{++}+N_{--}-N_{+-}-N_{-+}}{N_{++}+N_{--}+N_{+-}+N_{-+}},
\end{equation}
where $N_{++}$, $N_{--}$, $N_{+-}$ and $N_{-+}$ are the coincident
counts between Alice and Bob with the actual settings of $(\theta_A,
\theta_B)$, $(\theta_A+\pi/2, \theta_B+\pi/2)$, $(\theta_A,
\theta_B+\pi/2)$ and $(\theta_A+\pi/2, \theta_B)$, respectively. In
the CHSH inequality, parameter $S$ is defined as
\begin{equation}
S=|E(\theta_A, \theta_B)-E(\theta_A, \theta_B')+E(\theta_A',
\theta_B)+E(\theta_A', \theta_B')|.
\end{equation}
In the local realistic view, no matter what angles $\theta_A$ and
$\theta_B$ are set to, parameter $S$ should be $S\leqslant 2$.
However, in the view of quantum mechanics, $S$ will reach the
maximal value $2\sqrt2 \approx 2.828$ when the polarization angles
are set to $(\theta_A, \theta_A', \theta_B, \theta_B')=(0^\circ,
45^\circ, 22.5^\circ, 67.5^\circ)$. The observed values of the
correlation functions are shown in Fig.4, resulting in
$S=2.54\pm0.02$, which violates Bell's inequality by 27 standard
deviations. This clearly confirms the quantum nature of the
entanglement state.

\begin{figure}[t]
\includegraphics[width=8.4cm]{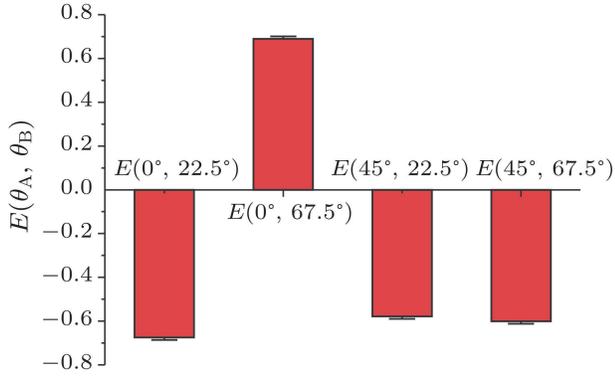}
\caption{Experimentally measured correlation function $E$ required
for CHSH inequality. Every measurement of $E$ is finished within 3
s. These data yield the Bell parameter $S=2.54\pm0.02$ which is in
confliction with local realism by over 27 standard deviations. The
error bars denote one standard deviation, deduced from propagated
Poissonian counting statistics of the raw detection events.}
\label{bell}
\end{figure}

Ultra high stability has been a major challenge for our experiment.
The interferometer's geometry has carefully been designed in a way
which minimizes unwanted changes in path length due to disturbances.
A glass chamber is constructed to protect the interferometer from
unwanted temperature drift and mechanical vibration. A chamber is
designed to allow control for all necessary angular settings from
the outside for all four subexperiments of the CHSH-inequality
measurement, after which we performed a measurement in Alice and
Bob's site with $\left\vert +\right\rangle/\left\vert
-\right\rangle$ basis to confirm that we are still in the
$\left\vert \Psi^{-}\right\rangle_{AB}$ state. Compactness of the
setup, the Home-built chamber and temperature control ($\pm 1^{\circ
}$) enable keeping the phase stable for several minutes which is
sufficient for our inequality measurement. Detection events are
registered by a self-developed constant fraction discriminator
(output signal width 2ns) and self-developed counting module in
combination with NIM-electronics for the logic.

In conclusion, our work supports that creation of entanglement
between two independent photons is possible, even if there is no
direct local interaction between the involved photons. The possible
choices of the paths the photons take can be seen as the underlying
cause for the observed non-local generation of entanglement. It is
worth to remark that, together with synchronized independent
narrow-band single photons\cite{Yuan2007} (typically have coherent
length 5m) and active phase stabilization, our experiment can be
extend to very long distance. The observed nonlocal two-photon
interference and generation of entanglement are deeply rooted in the
fundamentals of quantum mechanics and the question for hidden
variables. Besides the interest to fundamental physics, the
developed method has also practical applications, for example,
non-locally generated entanglement can be directly utilized as
essential resource for quantum-key distribution. If substituting
independent photons for polarization entangled photons, we can
readily prepare the two-photon four-dimensional entanglement and
cluster state which can be employed to perform the test of
`all-versus nothing' quantum nonlocality
\cite{Chen2003,Yang2005,Diao2008} and realize one-way quantum
computation.\cite{Walther2005,Chen2007} Our results may find
applications in all-optical quantum information processing.

\end{document}